\documentclass[prl,twocolumn,superscriptaddress,showpacs]{revtex4}
\usepackage{amsmath}
\usepackage{bm}
\usepackage{epsfig}
\usepackage{color}
\usepackage{float}
\usepackage{bm}
\usepackage{graphicx}
\renewcommand{\phi}{\varphi}

\newcommand{\revise}[1]{{\color{black} {#1}}}

\begin{document}

\title{Equilibrium sampling of hard spheres up to the jamming density 
and beyond}

\author{Ludovic Berthier}

\author{Daniele Coslovich}

\author{Andrea Ninarello}

\affiliation{Laboratoire Charles Coulomb, 
UMR 5221 CNRS-Universit\'e de Montpellier, Montpellier, France}

\author{Misaki Ozawa}

\affiliation{Laboratoire Charles Coulomb, 
UMR 5221 CNRS-Universit\'e de Montpellier, Montpellier, France}

\affiliation{Department of Physics, Nagoya University, Nagoya, Japan}

\newcommand{\rlj}[1]{{\color{blue}#1}}
\newcommand{\todo}[1]{{\color{red}\textbf{#1}}}

\date{\today}

\pacs{05.10.-a, 05.20.Jj, 64.70.Q-}




\begin{abstract}
We implement and optimize a particle-swap Monte-Carlo 
algorithm that allows us to thermalize a polydisperse system of 
hard spheres up to unprecedentedly-large volume fractions,
where \revise{previous} algorithms and experiments fail to equilibrate.
We show that no glass singularity intervenes before 
the jamming density, which we independently determine through
two distinct non-equilibrium protocols. We demonstrate 
that equilibrium fluid and non-equilibrium jammed states can have 
the same density, showing that
the jamming transition cannot be the end-point of the fluid branch.
\end{abstract} 

\maketitle

We clarify the behavior 
of non-crystalline states of hard spheres at 
very large densities where both a glass transition
(in colloidal systems) and a jamming transition (in non-Brownian systems) 
are observed~\cite{Liu_Nagel_1998,Parisi_Zamponi_2010,Berthier_Biroli_Bouchaud_Cipelletti_Saarloos_2011}. 
Glass and jamming transitions are usually studied through distinct protocols, 
and understanding the relation between these two broad classes of 
phase transformations and the resulting amorphous arrested states
is an important research 
goal~\cite{Liu_Nagel_1998,Parisi_Zamponi_2010,Berthier_Biroli_Bouchaud_Cipelletti_Saarloos_2011,Pica10,Ikeda_Berthier_Sollich_2012}. 
These questions 
impact a wide range of fields, from the rheological properties of soft 
materials to optimization problems in computer 
science~\cite{Liu_Nagel_1998,mezard}.

Let us first consider 
Brownian hard spheres. When size polydispersity
is introduced, crystallization can be prevented and the 
thermodynamic properties of the fluid 
studied at increasing density until a glass transition
takes place, where particle diffusivity becomes very small~\cite{Pusey87}. 
Upon further compression, the 
pressure of the glass increases until a jamming transition occurs, 
where particles come at close contact and the pressure 
diverges~\cite{Rintoul96}.
Because the laboratory glass transition arises from using 
a finite observation timescale, two scenarios were proposed 
to describe the hypothetical situation where thermalization 
is no longer an issue~\cite{Berthier_Witten_2009}. 
A first possibility is that slower compressions
reveal an ideal glass transition density, $\phi_{0}$,
above which the equilibrium state is a glass, not a 
fluid~\cite{Parisi_Zamponi_2010}.
Jamming would then be observed upon further non-equilibrium 
compression of these glass 
states~\cite{Mari_Krzakala_Kurchan_2009}. 
Alternatively, it may be that slower compressions 
continuously shift the kinetic glass transition to higher densities. 
In this view, it is plausible that jamming 
becomes the end-point of the equilibrium fluid 
branch~\cite{Aste_Coniglio_2004,Kamien_Liu_2007}. 

\begin{figure}[b]
\psfig{file=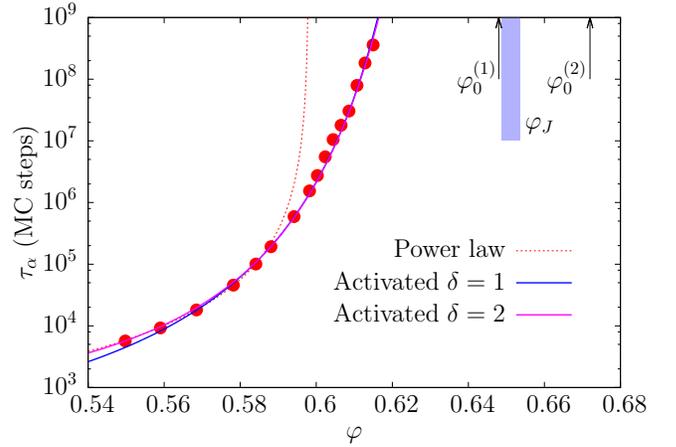,width=8.5cm}
\caption{Equilibrium relaxation timescale of polydisperse ($\Delta=23\%$) 
hard spheres from conventional Monte-Carlo simulations. The power law fit 
provides an estimate of the location of the
avoided mode-coupling transition, $\phi_{\rm mct} \approx 0.598$;
activated relaxation fits, $\exp(B/(\phi_0-\phi)^\delta)$,
yield possible locations for the extrapolation 
of a diverging timescale, $\phi_0^{(1)} \approx 0.648$ when $\delta=1$
and $\phi_0^{(2)} \approx 0.672$ for $\delta=2$. 
The range of jamming densities {$0.6487 < \phi_J < 0.6534$} 
using non-equilibrium 
protocols (shaded area) overlaps with the range of extrapolated dynamical 
divergences.}
\label{fig1}
\end{figure}

Distinguishing between these two scenarios by direct numerical
measurements is challenging. For a well-studied binary mixture of hard spheres,
for instance, thermalization can be achieved up to $\phi_{\rm max} 
\approx 0.60$~\cite{Berthier_Witten_2009,Brambilla_ElMasri_Pierno_Berthier_Cipelletti_Petekidis_Schofield_2009}.
The location of the glass transition must be extrapolated 
using empirical fits based on activated relaxation. 
Values in the range $\phi_{0} = 0.615 - 0.635$ were 
obtained~\cite{Berthier_Witten_2009}, 
depending on the fitting function. Fitting the relaxation times to a power law 
yields $\phi_{\rm mct} \approx  0.59 < \phi_{\rm max}$, 
so that the associated mode-coupling 
transition~\cite{Gotze_2009} corresponds
to an avoided singularity. For the same 
system, jamming transitions were located in the range
$\phi_J = 0.648 - 0.662$ depending on the chosen 
protocol~\cite{OHern_Langer_Liu_Nagel_2002,Chaudhuri_Berthier_Sastry_2010,Ozawa_Kuroiwa_Ikeda_Miyazaki_2012}. 
The relation between glass and jamming transitions 
is left unresolved, as thermalization stops long before 
any of the singularities can be crossed, $\phi_{\rm max} \ll 
\phi_{0}, \phi_J$, and because estimates of $\phi_0$ and 
$\phi_J$ are too close to favor any of the above scenarios. 
Similar inconclusive results are
obtained for polydisperse hard spheres with 
a continuous particle size distribution, see Fig.~\ref{fig1}.
Our standard Monte-Carlo simulations
yield slightly different values for the various critical 
densities, but again fall out of equilibrium too early to disentangle 
glass and jamming transitions.

Our main achievement is the numerical proof that
firm conclusions on these questions can be drawn by
implementing a more efficient Monte-Carlo sampling method.
The key enabling factor of our computational approach is 
the combination of swap Monte-Carlo moves to standard 
single-particle translations~\cite{pastore,pronk,Grigera_Parisi_2001},
\revise{which is a simple instance of a cluster move~\cite{santen,krauth}.}
Swap moves enhance thermalization in simple 
mixtures~\cite{Gutierrez_Karmakar_Pollack_Procaccia_2015}, but their efficiency 
is significantly higher in continuously polydisperse 
systems~\cite{dave,Fernandez_Mayor_Verrocchio_2007,ninarello_tobe}.
We show that this optimized sampling method allows to 
thermalize polydisperse hard spheres up to unprecedentedly-large densities. 
In particular, 
we achieve thermal equilibrium in a region of densities 
where the jamming transition  $\phi_J$
can also be located by accepted
methods~\cite{OHern_Langer_Liu_Nagel_2002,Berthier_Witten_2009,Xu_Blawzdziewicz_O'Hern_2005}. 
This directly shows, without extrapolating
any dynamical data, that point $J$ of 
jamming does not correspond to the end-point 
of the fluid equilibrium states. Our results also imply that the ideal 
glass transition, 
if it exists, occurs in our system
at an even larger density, $\phi_0 > \phi_J$. Conceptually, our results 
demonstrate that jamming and glass transitions 
explore distinct parts of the configuration space, respectively far from 
and close to equilibrium.

We study the canonical model of hard spheres in 
three dimensions. The pair interaction is zero
for non-overlapping particles, infinite otherwise. 
We use a continuous size polydispersity, where the particle
diameter is distributed according to $p(\sigma_{\rm m} \leq
\sigma \leq \sigma_{\rm M}) = A/\sigma^{-3}$, 
where $A$ is a normalization. Because of our enhanced 
Monte-Carlo sampling (see below), 
crystallization, which does not occur for ordinarily studied 
size polydispersities, $\Delta=\sqrt{\overline{\sigma^2} - 
\overline{\sigma}^2}/\overline{\sigma} \sim 8-14 \%$, and standard Monte-Carlo 
dynamics, is easily observed in our simulations. 
Therefore, to penetrate deeper 
the relevant glassy region and avoid crystal formation~\cite{bea}, 
we use a larger
polydispersity of $\Delta = 23 \%$, choosing $\sigma_{\rm m} / 
\sigma_{\rm M} = 0.4492$. We simulate systems composed of 
$N=300$, $1000$, and $8000$ particles 
in a cubic cell with periodic boundary conditions.
We present mainly data for $N=1000$, except where 
otherwise specified. 
Dynamical relaxation is recorded by measuring the self-intermediate 
scattering function $F_s(k,t)$ at a wave-vector $k$ corresponding to the 
first maximum of the structure factor. In constant 
volume simulations, we measure the 
pressure $P$ using the contact value of the pair correlation 
function~\cite{Santos05}. 
We also perform constant pressure simulations where $\phi$ 
fluctuates~\cite{FS01} \revise{\footnote{\revise{To adjust the pressure, 
we attempt compression - decompression moves at an average frequency
of one box move every $N$ translational particle move. The typical attempted
changes in the linear size of the box are about 3 \%.}}}.
The reduced pressure is 
$Z = P/ (\rho k_B T)$ where $\rho$ is the number density and 
$k_B T$ is the thermal energy. {Times are reported using 
standard Monte-Carlo timesteps~\cite{MCLJ}.

Starting with conventional Monte-Carlo dynamics with translational 
moves drawn over a cube of side $0.1 \overline{\sigma}$, 
we measure the equilibrium relaxation time $\tau_\alpha$ of the
system from the time-decay of the self-intermediate 
scattering function, $F_s(k,\tau_{\alpha})=1/e$, see Fig.~\ref{fig1}.
We observe an appreciable growth of about 5 orders of magnitude and
we can use standard functional forms to locate several 
characteristic densities for viscous slowdown. 
First, we locate the mode-coupling crossover using a power law,
$\tau_\alpha \sim (\phi_{\rm mct} - \phi)^{-\gamma}$. We get 
$\phi_{\rm mct} \approx 0.598$. As shown 
in Fig.~\ref{fig1}, we can thermalize the system 
beyond $\phi_{\rm mct}$ and conclude that this singularity
is avoided~\cite{santen,Brambilla_ElMasri_Pierno_Berthier_Cipelletti_Petekidis_Schofield_2009}. 
We then use an activated relaxation fit, $\tau_\alpha \sim \exp(B/(\phi_0 - \phi)^\delta)$, to 
extrapolate the location of the diverging 
relaxation time. We get 
$\phi_0^{(1)} \approx 0.648$ for $\delta=1$ and 
$\phi_0^{(2)} \approx 0.672$ using the exponent $\delta=2$ favored in earlier 
studies~\cite{Brambilla_ElMasri_Pierno_Berthier_Cipelletti_Petekidis_Schofield_2009,Berthier_Witten_2009}. The resulting range of 
$\phi_0$ proves that extrapolating $\phi_0$ from 
conventional simulations is a difficult exercise.

\begin{figure}
\psfig{file=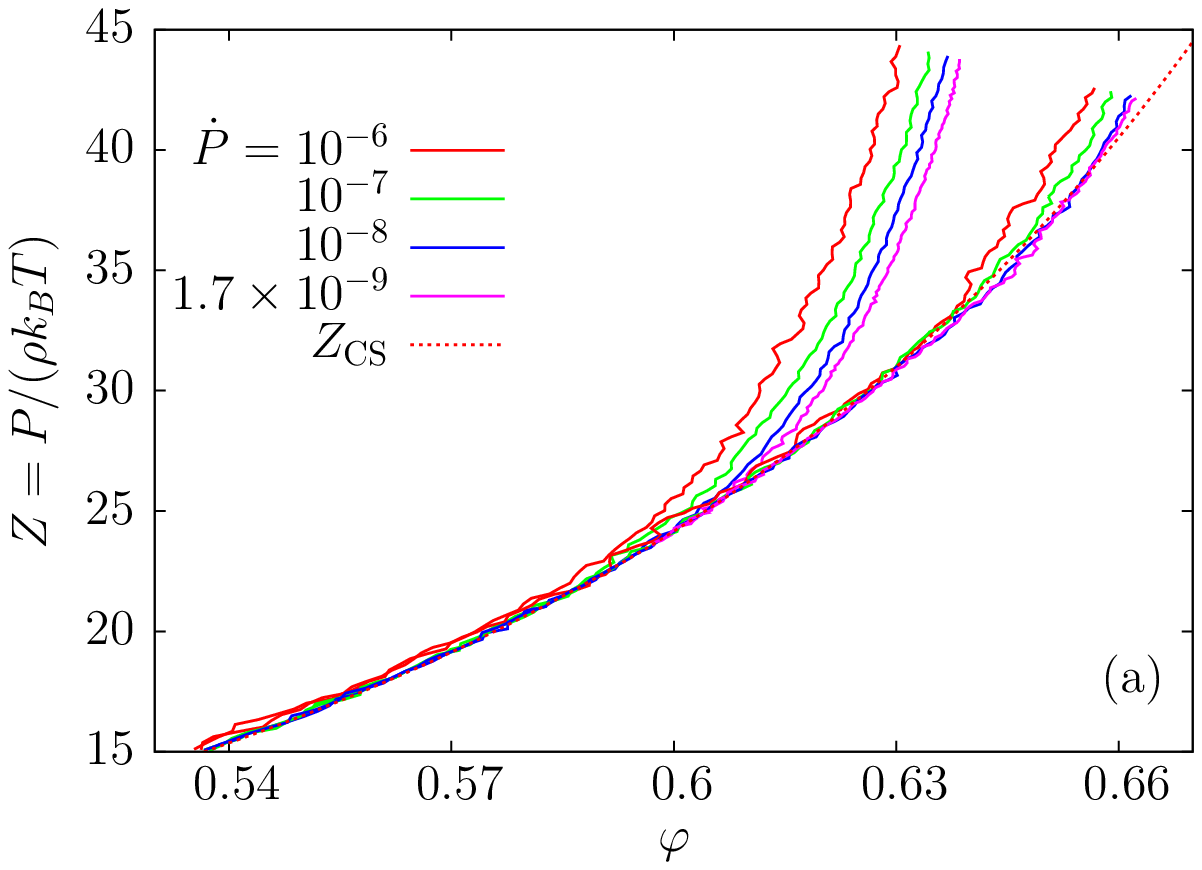,width=8.3cm}
\psfig{file=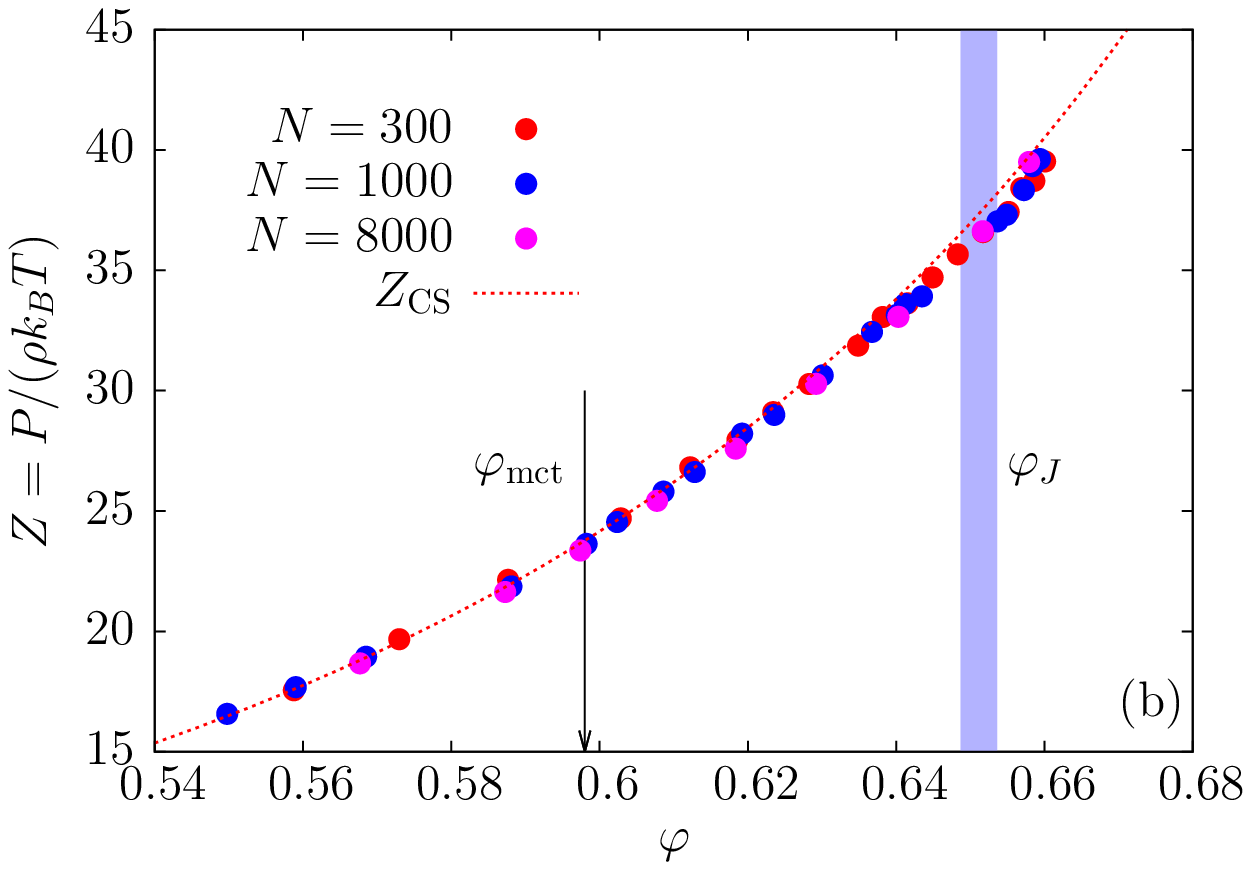,width=8.5cm}
\caption{(a) Reduced pressure $Z$ during compressions at finite 
rates $\dot P$ with (right curves) and without (left curves) swap.
The swap Monte-Carlo method falls out of equilibrium at much 
larger volume fractions than conventional sampling.
(b) Equilibrium measurements of the equation of state (symbols)
for different system sizes. Our equilibrium data go beyond the shaded area,
which represents the determined range of jamming densities. 
In both panels, the dashed line is the Carnahan-Stirling empirical 
expression~\cite{Boublik_1970,mansoori1971equilibrium}.}

\label{fig2}
\end{figure}

When $\phi > \phi_{\rm max} \approx 0.605$, 
standard Monte-Carlo simulations fall out of equilibrium. 
This is proved in Fig.~\ref{fig2}a where we report 
the evolution of $Z=Z(\phi)$ in 
simulations where we slowly increase the pressure 
at a finite rate $\dot P$ varying from  $\dot P = 10^{-6}$ down to 
$1.7 \times 10^{-9} $. Changing $\dot P$ over nearly 4 orders
of magnitude shows that thermalization is achieved up to increasingly
large densities, but $\phi_{\rm max}$ increases slowly with decreasing 
$\dot P$, with clear deviations from the fluid equation 
of state when $\phi > \phi_{\rm max}$. 

We now boost thermalization 
by implementing a swap algorithm~\cite{pastore,pronk,Grigera_Parisi_2001}.
With probability $\alpha$, we perform conventional Monte-Carlo moves
where particles are translated, and with probability $(1-\alpha)$
we perform a swap move where we pick two particles at random
and attempt to exchange their diameters. The swap is accepted if it 
does not create an overlap. By measuring the structural relaxation 
times with swap Monte-Carlo dynamics, we find that 
\revise{$\alpha \approx 0.80$} is the optimal
value for fast thermalization~\cite{ninarello_tobe}, with an 
efficiency that, remarkably, does not depend on system size.

This algorithm respects detailed balance and thus provides 
a correct sampling of phase space. That it also 
dramatically enhances the sampling efficiency is demonstrated in
Fig.~\ref{fig2}a where we compress the system using 
the same compression rates $\dot P$ as before, but now 
combining swap and translational moves. 
Whereas the new sampling method again falls out of equilibrium 
at large enough densities, we observe that
for $\phi < \phi^{\rm swap}_{\rm max} \approx 0.655$, 
$Z(\phi)$ does not depend on $\dot P$. The increase
of $\phi_{\rm max}$ from $0.605$ to $0.655$ is significant, as 
$\dot P$ should be slower by many orders of magnitude 
for the standard dynamics to perform as well as the swap method. 
The system shows no tendency to crystalline order, which 
would appear as a horizontal liquid-crystal coexistence plateau
in Fig.~\ref{fig2}a. 
We further checked the absence of crystalline order for $\Delta=23\%$ 
using standard bond-orientational order 
analysis~\cite{Steinhardt_Nelson_Ronchetti_1983}
\revise{and analysis of locally favored structures~\cite{coslovich_understanding_2007}.
We also checked that fractionation and size segregation do not occur by observing partial structure factors for higher and smaller particles.}
By contrast, we observed crystallization at large $\phi$ 
for lower values of $\Delta$.
Thus, we conclude from Fig.~\ref{fig2}a that our model is both robust 
against crystallization and easy to thermalize at extremely
large densities using optimized swap Monte-Carlo dynamics.
\revise{This represents a major methological progress for computer simulations 
of glassy materials.}

We now perform equilibrium measurements
of the pressure for each density where thermalization can be achieved.
By carefully checking that aging is absent  
in fully relaxing time correlation functions
(in particular density correlations)  
in long simulation runs of \revise{up to $10^{10}$ swap Monte-Carlo steps}, 
we confirmed
that thermalization is achieved for all $N$ up to $\phi^{\rm swap}_{\rm max} 
\approx 0.655$, directly establishing that no ergodicity breaking 
occurs below this point.
Equilibrium results for the equation of state 
$Z(\phi)$ are reported in Fig.~\ref{fig2}b for various 
systems sizes. 
We find that the empirical
Carnahan-Starling (CS) equation of state for polydisperse hard 
spheres~\cite{Boublik_1970,mansoori1971equilibrium}, 
$Z_{\rm CS}$, describes our numerical data surprisingly well up to 
the largest studied density. Closer inspection reveals
that small deviations from $Z_{\rm CS}$ appear near $\phi \approx 0.63$ 
above which $Z < Z_{\rm CS}$. Earlier reports 
had instead found opposite deviations from the CS equation of 
state~\cite{Rintoul96,Hermes_Dijkstra_2010,hermes2010jamming,OB11}, 
probably due to insufficient thermalization, as we also 
observe when using finite compression rates. \revise{Also, we do not find
evidence for a singular density describing the  
equilibrium equation of state.}
  
From these equilibrium data, it is clear that
the extrapolated singularity at $\phi_0^{(1)} = 0.648$
(Fig.~\ref{fig1}) is not observed, which shows that
the activated relaxation fit 
with $\delta = 1$, akin to the Vogel-Fulcher-Tamman law for 
molecular glasses~\cite{angell},
must break down between 
$\phi_{\rm max} = 0.605$ and $\phi_0^{(1)} =0.648$. 
An ideal glass transition, if present, 
must occur above $\phi^{\rm swap}_{\rm max}=0.655$. Therefore, our second 
extrapolated value $\phi_0^{(2)}=0.672$ is not ruled out.

We now determine \revise{the location of the point $J$ of jamming} 
for our system. Following earlier works, we employed two distinct numerical 
protocols. In a first approach, we   
perform very rapid compressions of dilute hard sphere fluid configurations
using Monte-Carlo simulations without swap~\cite{Berthier_Witten_2009}. We 
impose a very large pressure, $P=10^6$, and measure the long-time limit
of the volume fraction. 
These fast compressions yield $\phi_J^{(1)} \approx 0.6487$. 
In an independent protocol, we use energy minimization procedures converting 
the hard sphere potential into soft harmonic repulsive 
springs~\cite{OHern_Langer_Liu_Nagel_2002,Xu_Blawzdziewicz_O'Hern_2005}, 
which defines `point $J$'~\cite{OHern_Langer_Liu_Nagel_2002}. 
We iteratively compress or expand the system starting from 
a dilute random system distribution until particles are precisely 
at contact~\cite{Xu_Blawzdziewicz_O'Hern_2005}.
We obtain $\phi_J^{(2)} \approx 0.6534$. These two values are compatible
with earlier determinations of the jamming transition for 
polydisperse systems~\cite{Desmond_Weeks_2014}. 
\revise{Note that slower protocols 
would jam at larger packing fractions~\cite{Chaudhuri_Berthier_Sastry_2010}. }
While jammed configurations produced by the 
energy minimization protocol are always isostatic, 
those obtained using very rapid Monte-Carlo compressions are 
slightly hypostatic~\cite{Donev_Torquato_Stillinger_2005}. This 
might explain the small difference between $\phi_J^{(1)}$ 
and $\phi_J^{(2)}$~\cite{Chaudhuri_Berthier_Sastry_2010}.

In Fig.~\ref{fig2}b, we show that the equilibrium 
equation of state of the fluid can be measured up to $\phi_J$
{\it and beyond}, demonstrating that 
$\phi_J$ cannot be the end-point of the fluid branch.  
These data indicate therefore 
that $\phi_0$, if it exists, must actually be larger than $\phi_J$. 
If the jamming transition does not control the large-pressure 
behavior of the equilibrium fluid, its potential connection to glassy 
dynamics is then considerably weakened. 
Fundamentally, the glass/jamming decoupling
exposed by our measurements occurs because  
jamming is observed using protocols that are very far from thermal equilibrium. 
Instead, in the presence of thermal fluctuations, the configurational space 
is sampled with the Boltzmann probability distribution.
Distinct sets of configurations are thus explored 
in thermal and athermal protocols.

\begin{figure}
\psfig{file=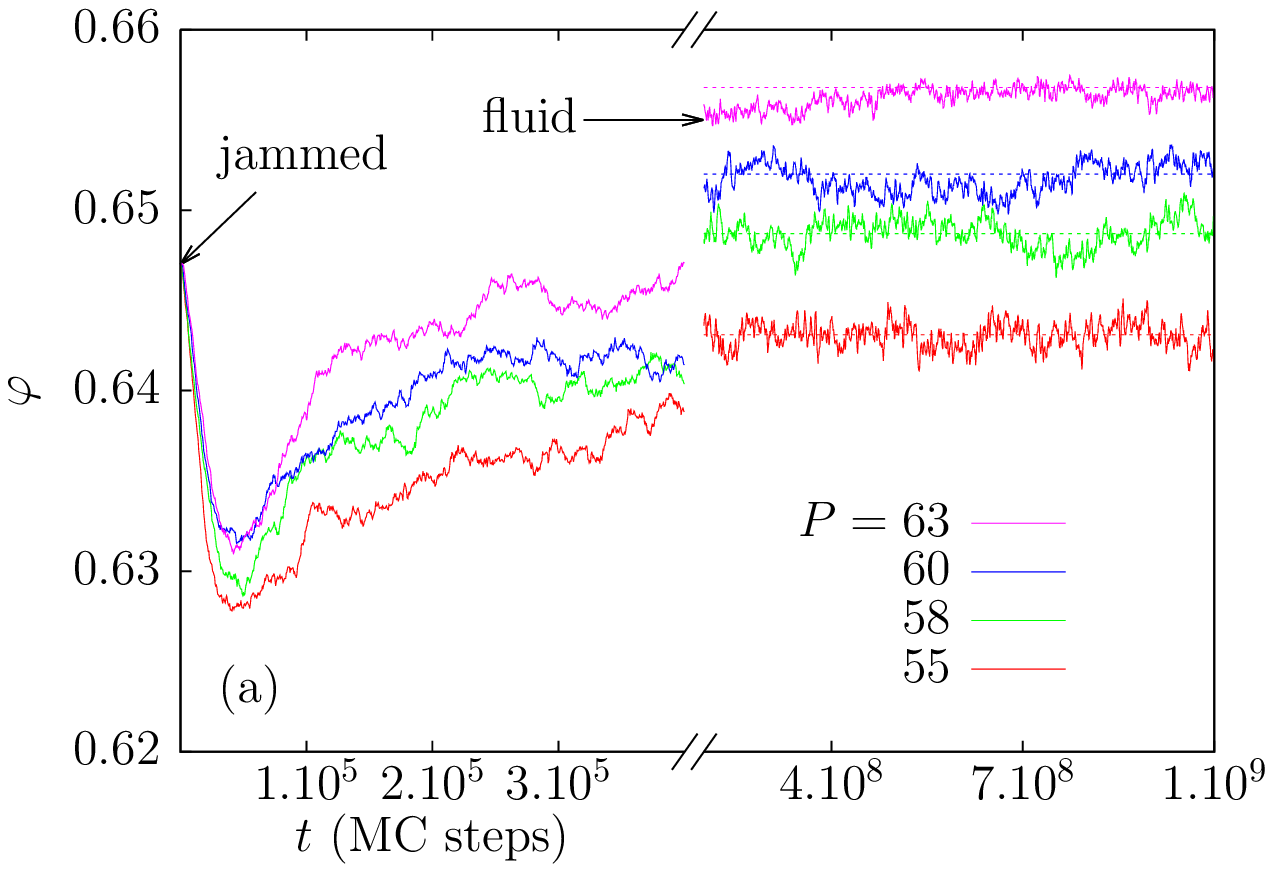,width=8.5cm}
\psfig{file=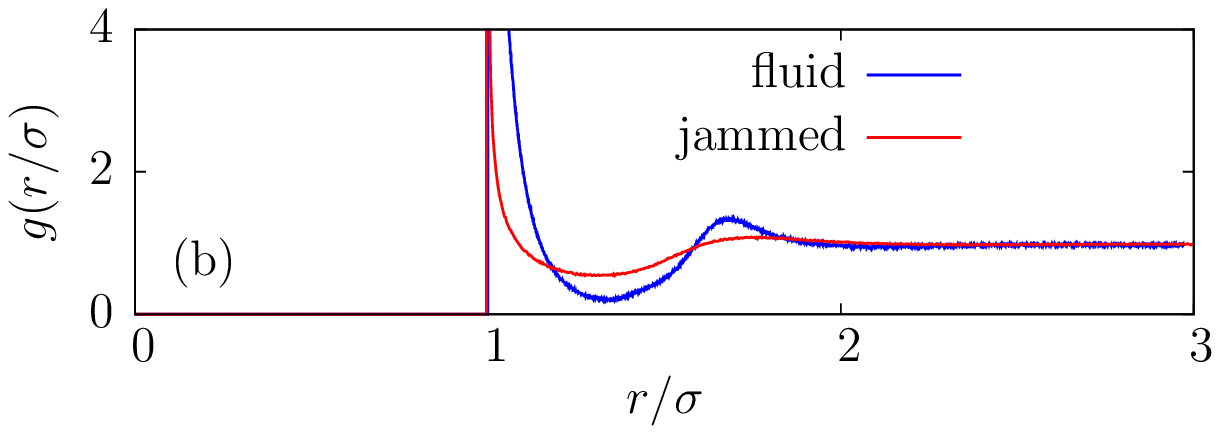,width=8.cm}
\caption{(a) Memory effect in hard spheres: 
Time evolution of the volume fraction starting at $t=0$ from a jammed 
configuration in constant pressure swap Monte-Carlo simulations.
A non-monotonic evolution of the density is observed, 
showing that jammed and fluid 
states are structurally distinct. At long times (note the break in 
the time axis), steady state 
fluctuations are observed both below and above $\phi_J$ around the
value (dashed lines) obtained from the equation of 
state. (b) Pair distribution functions of rescaled distances
$r_{ij}/[(\sigma_i+\sigma_j)/2]$ are distinct for fluid and 
jammed states near $\phi=0.6534$.}
\label{fig3}
\end{figure}

To confirm both this fundamental difference
and to reinforce our conclusion that fluid and jammed states
may exist over a similar density range, we perform
the memory experiment shown in Fig.~\ref{fig3}(a). 
We start simulations using the nearly jammed configurations
created by fast Monte-Carlo compressions of 
dilute fluid states. The pressure in these states 
is very large, $P = 10^6$. 
At time $t=0$, we start a constant pressure Monte-Carlo simulation 
involving particle swaps, for a range of pressures $P$ chosen 
such that the corresponding equilibrium volume fractions 
lie in the vicinity of $\phi_J$. Therefore, at time $t=0^+$, 
the volume fraction has a value which is already very close to its equilibrium
value. If jammed and fluid states were structurally close to one another, 
the density should exhibit a mild time dependence, 
having similar values at small and large times. 
The results in Fig.~\ref{fig3}(a) instead demonstrate a more complex 
time dependence with a pronounced 
non-monotonic behavior. Readers familiar with the 
physics of disordered materials will recognize this protocol 
as a memory (or Kovacs) effect~\cite{Kovacs_1963}. 
Similar non-monotonic behaviors were observed 
in a variety of glassy systems,
from polymer glasses to granular media and 
spin glasses~\cite{angell,leheny,levelut,josserand,BB02,mossa}.
 
As soon as the pressure is set to a finite value, the system 
undergoes a rapid expansion from the jammed initial configuration,
accompanied by little particle diffusion. 
Under the influence of the \revise{swap} Monte-Carlo dynamics, particles 
diffuse and the system starts to thermally explore 
fluid configurations in order to reach equilibrium at long times.
During this aging process, the volume fraction slowly increases,
which explains the non-monotonic time dependence.  
At very large times, the density reaches its 
steady state equilibrium value, which 
depends on the applied pressure through the equilibrium equation 
of state. The time series in Fig.~\ref{fig3} 
indicate that stationary fluid states exist both below, at, and 
above $\phi_J$, whereas the non-monotonic time dependence of the 
volume fraction demonstrates that typical jammed and 
high density fluid configurations are structurally very distinct,
as testified by the pair correlation functions 
shown in Fig.~\ref{fig3}(b). \revise{The structure of fluid and 
jammed states differs not only near contact (where the jammed 
$g(r)$ is highly singular), but also further away from contact.}

In conclusion, we presented numerical measurements 
of the fluid equation of state of hard spheres in a part of the phase 
diagram that was so far unaccessible to both numerical simulations 
and experiments. Our results rule out the possibility that the  
jamming transition represents the end-point of the fluid 
equation of state~\cite{Kamien_Liu_2007,Aste_Coniglio_2004}, 
and suggest that a glass transition, 
if it exists, is logically disconnected from the 
jamming transition. This situation
contrasts with both leading scenarios discussed in the introduction,
although it fits naturally in theories where jamming densities 
span a finite range of protocol-dependent 
values~\cite{Mari_Krzakala_Kurchan_2009,Parisi_Zamponi_2010}.
A similar scenario is found in quasi-one dimensional channels, 
where equilibrium fluid and jammed states also coexist over a density
range~\cite{ashwin}.

The efficiency of the optimized Monte-Carlo sampling developed here, 
combined with a sufficient degree of particle size 
polydispersity~\footnote{Whereas our main conclusion about the 
interrelation between glass and jamming transitions should hold
generally for arbitrary particle size distribution, continuous 
or discrete, the possibility to develop an efficient 
swap Monte-Carlo dynamics allowing the exploration of phase space 
at large density will depend on the chosen 
distribution~\cite{ninarello_tobe}.}, 
paves the way 
for a set of novel studies of glassy states in amorphous materials.
We expect this approach to be fruitful in elucidating a number 
of outstanding aspects of the glass problem, such as the existence of the 
Gardner transition in finite dimensional systems~\cite{CKPUZ14,preprint}, 
the growth of static point-to-set 
correlations~\cite{BBCGV08,PhysRevLett.108.225506,sho}
and locally-favored structures~\cite{paddy}, measurements of 
configurational entropy~\cite{BC14}, and the physics 
of ultrastable glasses~\cite{ultra1}.

\acknowledgments

We thank J. Kurchan for discussions.
The research leading to these results has received funding
from the European Research Council under the European Union's Seventh
Framework Programme (FP7/2007-2013) / ERC Grant agreement No 306845.
M. O. acknowledges the financial support by Grant-in-Aid 
for Japan Society for the Promotion of Science Fellows (26.1878).

\bibliography{paper}

\end{document}